\documentclass[12pt]{article}

\usepackage{times,amsfonts,epsfig,psfrag}
\usepackage[latin1]{inputenc}
\usepackage{amsfonts}
\usepackage{amsmath}
\usepackage{amssymb}
\usepackage{amsthm}
\usepackage[american]{babel}
\usepackage[T1]{fontenc}
\usepackage{slashed}
\usepackage{dsfont}
\usepackage{wick}
\usepackage{array}
\usepackage{epsfig}
\usepackage{amssymb}
\usepackage{graphics,graphpap}

\newcommand{\qbar} {\overline{q}}

\newcommand{\bra}[1]{\left\langle #1 \right|} 
\newcommand{\ket}[1]{\left| #1 \right\rangle}

\newcommand{\nn}{\nonumber}  
\newcommand{\ubar}{\overline{u}}
\newcommand{\cbar}{\overline{c}}
\newcommand{\dbar}{\overline{d}}

\newcommand {\qcond} {\left\langle \qbar q\right\rangle} 
\newcommand{\Da}{\mathcal{D \underline{\alpha}} }

\allowdisplaybreaks

\usepackage{array}
\usepackage{epsfig}
\usepackage{amssymb}
\usepackage{graphics,graphpap}

\begin{document}

\begin{titlepage}
\vskip1.5cm
\begin{center}
{\Large \bf \boldmath
Determination of the magnetic susceptibility of the quark condensate using radiative heavy meson decays}
\vskip1.3cm 

{\sc J. \: Rohrwild}
\\[0.5cm]
\vspace*{0.1cm}{\it
Institut f\"ur Theoretische Physik, Universit\"at Regensburg,
\\D-93040 Regensburg, Germany} 
\\[1.0cm]

\vspace{0.6cm}
\bigskip
\centerline{\large \em \today}
\bigskip

\vskip3cm

{\large\bf Abstract\\[10pt]} \parbox[t]{\textwidth}{We use a light-cone sum rule (LCSR) analysis of the branching ratios of radiative meson decays to contrain the value of the magnetic susceptibility of the quark condensate $\chi(\mu)$. For the first time, we use a complete set of three-particle distribution amplitudes that enables us to give a consistent prediction for the branching ratios. Furthermore we will make use of a very recent update of several non-perturbative parameters. Our final result for $\chi(\mu= 1\;{\rm GeV})= 2.85 \pm 0.5\; {\rm GeV}^{-2}$ (assuming asymptotic wave functions) agrees with the currently used value of $3.15 \pm 0.3 \;{\rm GeV}^{-2}$.}

\end{center}
\end{titlepage}

\newpage
\setcounter{equation}{0}

\section{Introduction}
\addtocounter{page}{0}

The term ``magnetic susceptibility of the quark condensate'', $\chi(\mu)$, was coined as early as 1984 in a fundamental work by Ioffe and Smilga \cite{Ioffe:1983ju} on the magnetic properties of the QCD vacuum. The denotation is due to a very intuitive picture: the value of the non-perturbative parameter $\chi$ determines the strength of the reaction of the vacuum with respect to an applied external electromagnetic field, analogous to what is understood by magnetic susceptibility in solid states physics.

The magnetic susceptibility\footnote{As we only deal with one kind of ``magnetic susceptibility'', we will usually drop the reference to the quark condensate.} is an important input parameter in various calculations of processes involving real photons, such as dijet photo-production \cite{Ivanov}, radiative decays \cite{radiative} or the anomalous magnetic moment of muons \cite{Czarnecki:2002nt}.
It is known that $\chi$ is in fact surprisingly big compared to what one might assume as its ``natural'' scale, hence, it often provides the dominant contribution to the calculation, see \cite{sumrules} for examples. Therefore, while the determination of the magnetic susceptibility may provide additional insights into properties of the QCD vacuum, a task worthy in its own right, a precise knowledge of the value of $\chi(\mu)$ is important for its application of QCD calculations. 

Thus, various theoretical approaches to determine the magnetic susceptibility have been suggested. To the best of our knowledge, the first work regarding this topic was published in 1983 \cite{Balitsky:1983xk} and made  use of the so-called vector-dominance approximation. Others include consistency conditions of QCD sum rule predictions \cite{Ioffe:1983ju} and local  duality approaches \cite{Belyaev:1984ic,Balitsky:1985aq}. The value of $\chi(\mu=1{\rm GeV})=3.15 \pm 0.3 \;{\rm GeV}^{-2}$, which is currently used in most QCD sum rule \cite{SVZ} calculations was obtained by a combination of a local duality approach and a QCD sum rule for $\chi$ \cite{Ball:2002ps}. However, in \cite{Vainshtein:2002nv} the possibility of a larger value is considered using a rather elegant argument similar to the famous Gell-Mann--Oakes--Renner (GMOR) relation. An approach based on the instanton liquid model can be found in \cite{Petrov:1998kg,Kim:2004hd,Dorokhov:2005pg}.

In this work a different way to determine the value of $\chi$ is employed. We fit a $\chi$-dependent sum rule to an experimental value. We will argue (in Sec.\ref{2}) that radiative heavy meson decays and especially the $D^* \to D \gamma$ decay and the corresponding branching ratio are suited for such a procedure. Heavy meson decays have been studied numerously and successfully with sum-rule-based approaches, see e.g. \cite{HMD}. In \cite{Aliev:1995wi} the  different $D^* \to D \gamma$ and $B^* \to B \gamma$ decays have been examined using a light-cone sum rule approach based on the background field method, that was advocated in \cite{Balitsky:1989ry}. While we follow the general spirit of this calculation, we use a more recent complete set of photon distribution amplitudes up to twist 4 \cite{Ball:2002ps}, including some small contributions that where neglected in \cite{Aliev:1995wi}. We will also make use of a recently published update on some non-perturbative parameters \cite{Ball:2007zt}. As a detailed calculation of the various $D^* \to D \pi$ decays was presented in \cite{Belyaev:1994zk} and the $D^*$ decays only into $D \pi$ and $D\gamma$, we are able to determine the $\chi$-dependence of the corresponding branching ratios up to and including twist 4. 

The paper is organized as follows. In Chapter \ref{2} we will calculate the $D^{*+}\to D^{*+} \gamma$ decay constant $g_{D^*D\gamma}$ using LCSRs. The next Chapter deals with the branching ratios and how to extract the value of $\chi$. Chapter \ref{4} contains the conclusions.

\section{LCSRs for the $D^* \to D \gamma$ decay}
\label{2}

As already mentioned in the introduction, if one wants to find $\chi$ by fitting a sum rule to experiment, it is important to select a process the exhibits a strong dependence on $\chi$. This unfortunately rules out some well known classical problems like the magnetic moments of nucleons, as sum rule calculations \cite{sumrules} show large contributions due to higher twist terms, most notably twist 4, and only a moderate dependence on $\chi$. 
A suitable process should thus fulfill several conditions:
\begin{itemize}
\item in order to archive a better suppression of higher twist terms it is advantageous to consider processes that involve a propagating heavy quark, which will induce a $\frac{1}{m_q}$--suppression of twist-4 contributions
\item existing experimental data should be accurate enough to support a fit
\item the relevant sum rule should not contain any theoretical difficulties, e.g. sum rules for transition between states of different mass are known to require an adjustment of the Borel parameters to allow for the mass gap, which would possibly entail additional uncertainties.
\end{itemize}
This does not single out any particular process, however our choice, the $D^* \to D \gamma$ decay, seems to fulfill the criteria rather well \footnote{In fact, at first glance the decay $B^* \to B \gamma$ is even more appealing. Unfortunately the $B^*$ is too heavy to be produced in B-factories running at the $\Upsilon (4 s)$ resonance. Furthermore, due to kinematics the process $B^{*} \to B \pi$ is only possible as a virtual subprocess, therefore the branching ratio is not a good observable for our purpose.}. It is also advantageous that the decay $D^* \to D \pi$ has been analyzed using LCSRs \cite{Belyaev:1994zk}, thus allowing us to consider sum rules for the branching ratios, which have been measured rather precisely \cite{Bartelt:1997yu}. 

In the remainder of this Chapter we will examine the $D^* \to D \gamma$ transition. In Ref.\cite{Aliev:1995wi} this was already calculated using LCSRs in conjunction with photon distribution amplitudes. Our calculation will use, for the first time, a complete set of twist 4 DAs and updated parameters.
As the only difference between the $D^{*0}$ and the $D^{*+}$ is the charge of the light quark, we will focus on the case of the $D^+$ \footnote{To avoid unnecessary indices, we will usually abstain from using the $^+$.}.

\subsection{Definitions}

\begin{figure}
\begin{center}
\epsfig{figure=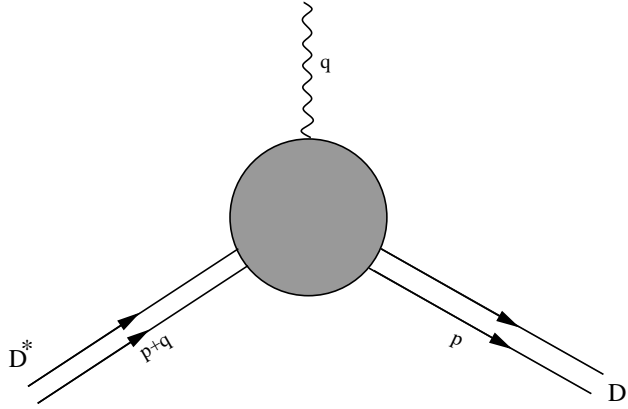,width=0.5\textwidth}
\end{center}
\caption{Radiative decays of a $D^*$ with momentum $p$ into a $D$ with momentum $p+q$. \label{DD}}
\end{figure}
The transition matrix element  
\begin{align}
\label{TME}
 \bra{D^{*}(p,\lambda)}j^{\mu}_{em}(0)\ket{D(p+q)}=\varepsilon^{\mu \nu \alpha \beta } p_{\nu} \epsilon_{(\lambda)\alpha} q_{\beta} g_{D^*D\gamma}(Q^2)	
 \mbox{ }
\end{align}
can be parametrized by introducing the transition amplitude $g_{D^*D\gamma}$. Here $\epsilon^{(\lambda)\beta}$ is the 4-polarization vector of the $D^{*}$. The decay width only depends on $g_{D^*D\gamma}(0)$, therefore, we only have to take real photons into account. Henceforth, $e^{(\lambda)}_{\mu}$ is the four-polarization vector of the emitted photon, $q \cdot e^{(\lambda)}=0$. 

The decay as shown in Fig.\ref{DD} can conveniently be described by the correlation function
\begin{align}
\label{CF}
\Pi^{\mu\nu} \left(p,q \right) = i^2  \int\! d^4 x  \int \!d^4 y \;e^{i px + i qy} \bra{0} \mathcal{T} \lbrace \eta^{\mu}_{D^*} \! \left( x\right) j^{\nu}\left( y \right) \overline{\eta}_{D}\! \left( 0 \right) \rbrace  \ket{0} \mbox{ .}
\end{align}
Here 
\begin{align}
\label{EmCurrent}
j^{\mu}=e_d \dbar \gamma_{\mu} d + e_c \cbar \gamma_{\mu} c
\end {align}
is the electromagnetic current, with $e_d=-1/3$ and $e_c=2/3$ being the quark charges. The currents \cite{Balitsky:1989ry}
\begin{align}
\label{currentEta*}
\eta^{\mu}_{D^*}(x)=\ubar^{a}(x) \gamma^{\mu} c^a(x)\\
\label{currentEta}
\eta^{\mu}_{D^*}(x)=\ubar^{b}(x) i \gamma_5 c^b(x)
\end{align}
are used to generate a state with the quantum numbers of a $D^{*}$ or a $D$ meson, respectively. $a,\;\!b$ are color indices.
The coupling constants $f_{D^*}$ and $f_{D}$ of the currents \eqref{currentEta*} and \eqref{currentEta} to the corresponding states are defined as \cite{Balitsky:1989ry}
\begin{align}
\label{overlappEta*}
\bra{0} \eta^{\mu}_{D^*}(0)\ket{D^{*}(\lambda, p)}=&m_{D^*} f_{D^*} \epsilon^{(\lambda) \mu}\\
\label{overlappEta}
\bra{0} \eta_{D}(0)\ket{D(p)}=&\frac{m_{D^*}^2}{m_c} f_{D}\;\mbox{ .}
\end{align}
Hereafter, $m_{D^*}$ and $m_D$ is the mass of the $D^*$ and the $D$. $m_c$ is the mass of the charm quark.

It is advantageous to rewrite the correlator by introducing an electromagnetic background field of a plane wave
\begin{align}
\label{Fmunu}
F_{\mu \nu} = i\left(e^{(\lambda)}_{\nu} q_{\mu}-e^{(\lambda)}_{\mu} q_{\nu} \right) e^{iqx} \mbox{ .}
\end{align}
This allows us to write the following correlation function
\begin{align}
 \label{Dcorrelator}
 \Pi_{P}^{\mu \nu} \left(p,q \right) e^{\left( \lambda \right)}_ {\nu}= i \int\! d^4 x   \;e^{i px} \bra{0} \mathcal{T} \lbrace \eta^{\mu}_{D^*} \! \left( x\right) \overline{\eta}_{D}\! \left( 0 \right) \rbrace  \ket{0}_F \mbox{ .}
\end{align}
Here the subscript $_{F}$ indicates that the vacuum expectation value has to be evaluated in the background field $F_{\mu \nu}$. 
The correlation function in Eq.\eqref{CF} can then be reproduced by expanding Eq.\eqref{Dcorrelator} in powers of the background field and taking only the terms linear in $F_{\mu \nu}$ corresponds to the single photon emission. A detailed analysis of the general procedure can be found in Ref.\cite{Ball:2002ps}, while an excellent review on the background field method can be found in \cite{Novikov:1983gd}. 

\subsection{Sum rule for $g_{D^*D\gamma}$}

Following the general strategy of QCD sum rules, the correlation function Eq.\eqref{CF} has to be evaluated in two different kinematic regions. On the one hand,  Eq.\eqref{Dcorrelator} will be dominated by the decay $D^* \to D \gamma$, if $p^2 \approx m_{D^*}^{2}$ and $(p+q)^2 \approx m_D^2$. Then again, in the kinematic region where $p^2 \ll 0$ and $(p+q)^2 \ll 0$ an expansion in terms of photon distribution amplitudes of increasing twist is valid.

On the hadronic level Eq.\eqref{CF} can then by written as
\begin{align}
\label{hadronic}
\Pi_{P}^{\mu \nu} \left(p,q \right) e^{\left( \lambda \right)}_ {\nu}= \frac{f_{D^*}f_{D} m_{D}^2 m_{D^*}g_{D^*D\gamma}}{m_c \left(m_{D^*}^2-p_1^2\right) \left( m_D^2-p_2^2 \right)} \varepsilon^{\mu \alpha \beta \rho} p_{\alpha} q_{\beta} e^{(\lambda)_{\rho}}+\ldots
\end{align}
Where we made use of Eqs.(\ref{TME},\ref{overlappEta*},\ref{overlappEta}) and introduced the abbreviations $p_1=p$ and $p_2=p+q$. The dots represent contributions from excited states and the continuum. 

To calculate the correlation function in the deep Euclidean regime is is necessary to insert the explicit expressions for the currents $\eta^{\mu}_{D^*}$ and $\eta_{D}$ into Eq.\eqref{Dcorrelator}, yielding
\begin{align}
\Pi^{\mu \nu}(p,q) e^{(\lambda)}_{\nu}=i \int {\rm d}^4x e^{ipx} \bra{0} \mathcal{T}\left \lbrace   \dbar^a(x) \gamma^{\mu} c^{a}(x) \cbar^{b}(0)i \gamma_5 d^{b}(0)  \right \rbrace \ket{0}_F\;\mbox{ .}
\end{align}
As both massive and massless quarks propagate in the presents of gluonic and electromagnetic backgroundfields the following expressions\footnote{We tacitly assume Fock-Schwinger Gauge, which allows us to omit the path-ordered exponents.} have to be used after employing Wick's theorem \cite{Balitsky:1987bk}:
\begin{align}
\label{masslesspropagator}
\wick{1}{<1{d}(x) >1{\overline{d} }}(0)=&\frac{i\slashed{x}}{2 \pi^2 x^4}-\frac{ig}{16 \pi^2 x^2}\int_0^1 \! {\rm d}u \; \left \lbrace \ubar \slashed{x} \sigma_{\alpha \beta} + u  \sigma_{\alpha \beta} \slashed{x} \right \rbrace G^{\alpha \beta}(ux)  \nn
\\
&-\frac{ie_d }{16 \pi^2 x^2} \int_0^1 \!du\;  \left \lbrace \ubar \slashed{x} \sigma_{\alpha \beta}+u  \sigma_{\alpha \beta}\slashed{x} \right \rbrace F^{\alpha \beta}(ux)+\ldots \mbox{  ,}
\end{align}
and 
\begin{align}
\label{massivepropagator}
\wick{1}{<1{c}(x) >1{\overline{c} }}(0)=&\int\frac{{\rm d}^4k}{(2\pi)^4i} e^{-ikx} \frac{\slashed{k}+m}{m_c^2-k^2}\nn \\ &-i g \int\frac{{\rm d}^4k}{(2\pi)^4i} e^{-ikx}\int_0^1 {\rm d}u \left[ \frac{\slashed{k}+m_c}{2(m_c^2-k^2)^2}G^{\mu \nu}(ux)\sigma_{\mu \nu} +\frac{u x_{\mu}}{m_c^2-k^2}G^{\mu \nu}(ux)\gamma_{\nu} \right] \nn \\
&-i e_c \int \!\!\!\!\frac{{\rm d}^4k}{(2\pi)^4i} e^{-ikx}\int_0^1 \!\!\!\!{\rm d}u \left[ \frac{\slashed{k}+m_c}{2(m_c^2-k^2)^2}F^{\mu \nu}(ux)\sigma_{\mu \nu} +\frac{u x_{\mu}}{m_c^2-k^2}F^{\mu \nu}(ux)\gamma_{\nu} \right] +\ldots  \; \mbox{ .}
\end{align}
where we used the common abbreviation $\overline{u}=1-u$. $F_{\mu \nu}(x)$ is the electromagnetic field strength tensor defined in Eq.\eqref{Fmunu} and $G_{\mu \nu}=G_{\mu \nu}^{A} t^{A}$ is the gluon field strength tensor. The dots represent terms that will not give rise  to terms of twist 4 or lower. 

\begin{figure}[t]
\centering{\epsfig{figure=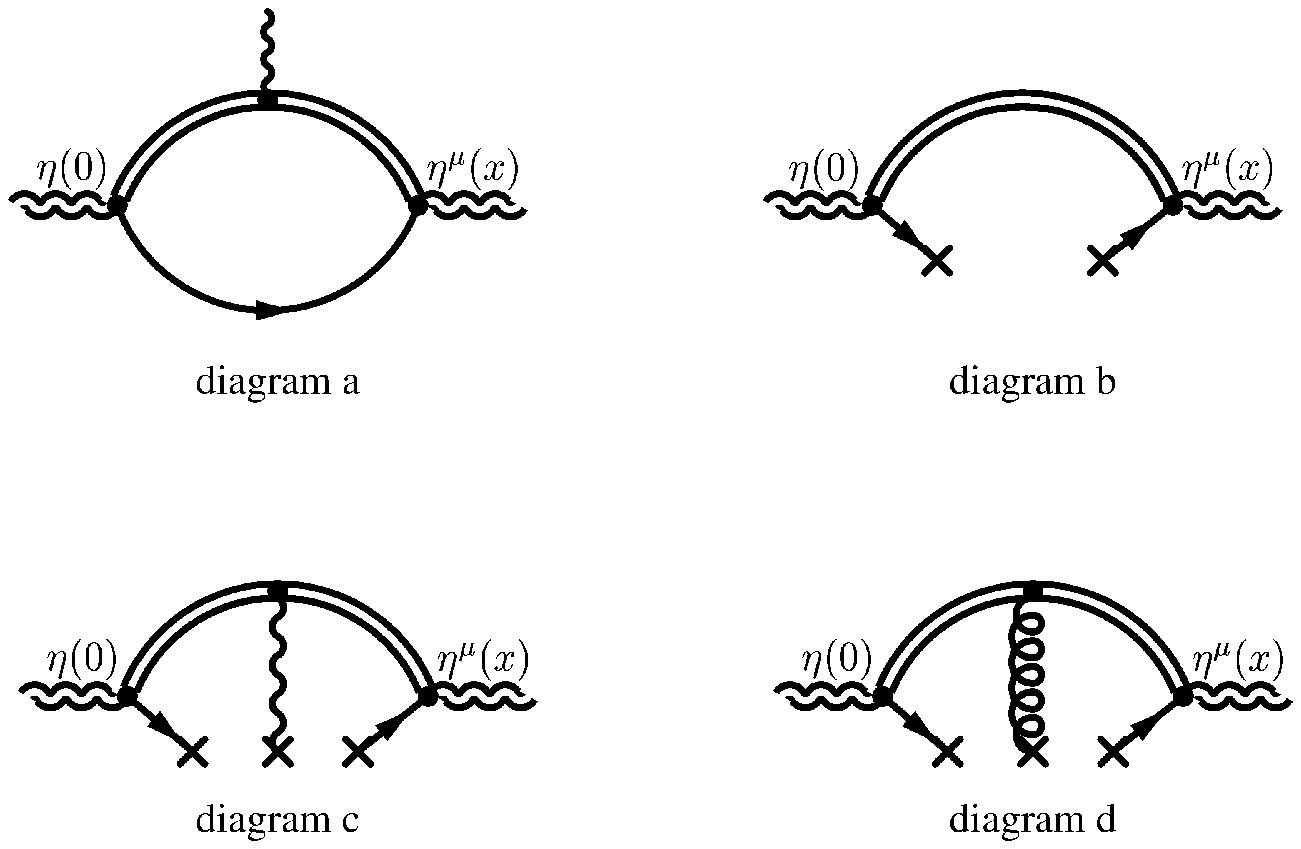,width=0.6\textwidth}}
\caption{Diagrams up to twist 4. The wiggled and the curled lines represent the coupling to the electromagnetic and gluonic background fields. The double lines denote the heavy quark and the single lines the  light quark. Photon distribution amlitudes are indicated by crosses. \label{diagrams}}
\end{figure}

Up to twist 4 there are only four relevant Feynman diagrams as depicted in Fig.\ref{diagrams} and their counterparts with exchanged heavy and light quarks. However, as the charm quark condensate vanishes identically, only the perturbative ``mirror diagram'' has to be considered. The calculation is then straight forward, although quite cumbersome, so that we will just give the final results for the different diagrams.\\
The sum of the two perturbative diagrams is given by
\begin{align}
\label{Ta}
T_{a}^{\mu}=-\frac{N_c}{4 \pi^2} \int_0^1 {\rm d} t \left[ t \frac{t e_c m_c} {t m_c^2-t \overline{t} (\ubar p_1^2 + up_2^2)} + t \frac{\overline{t} e_d m_c} {\overline{t} m_c^2 - t \overline{t} (\ubar p_1^2 + up_2^2)} \right] \varepsilon^{\mu \nu \alpha \beta} p_{\nu} q_{\alpha} e_{\beta}^{(\lambda)} \;\mbox{ .}
\end{align}
For the remaining diagrams one finds:
\begin{align}
\label{Tb}
T_{b}^{\mu}(p,q)=&\phantom{+}\bigg [e_d \qcond \int_0^1 {\rm d}u \left( \frac{\chi \varphi(u)}{m_c^2-(u p_1^2 + \ubar p_2^2)} \right.\nn \\ 
&\phantom{+ \bigg [e_d \qcond \int_0^1 {\rm d}u } \left.\;\;-\frac{\mathds{A}(u)}{4}\left( \frac{1}{(m_c^2-(u p_1^2 + \ubar p_2^2))^2} + 2 \frac{m_c^2}{(m_c^2-(u p_1^2 + \ubar p_2^2))^3} \right) \right) \nn \\
&\phantom{+ \bigg [}+e_d m_c \int_0^1 {\rm d}u \frac{f_{3\gamma}}{2(m_c^2-(u p_1^2 + \ubar p_2^2))^2}\psi^{A}(u) \bigg ]\cdot \varepsilon^{\mu \nu \alpha \beta} p_{\nu} q_{\alpha} e_{\beta}^{(\lambda)}
\\
\label{Tc}
T_{c}^{\mu}(p,q)=&\phantom{+} e_c \qcond \left[\int_0^1\! \!{\rm d}u \int\! \Da  \frac{1}{(m^2_c-(p+\alpha_uq)^2)^2} \left( \mathcal{S}_{\gamma}(\underline{\alpha}) - (1+2u) \mathcal{T}^{4}_{\gamma}(\underline{\alpha})\right)+2\mathds{P}\left[\mathcal{T}^{4}_{\gamma}(\underline{\alpha})\right]\right] \nn \\ & \qquad \cdot \varepsilon^{\mu \nu \alpha \beta} p_{\nu} q_{\alpha} e_{\beta}^{(\lambda)}
\\
\label{Td}
T_{d}^{\mu}(p,q)=&\phantom{+} e_d \qcond \left[\int_0^1 \! \! \! {\rm d}u \int\! \! \Da  \frac{1}{(m^2_c-(p+\alpha_uq)^2)^2} \left( \mathcal{S}(\underline{\alpha}) + (1+2u)(\mathcal{T}_3(\underline{\alpha})-\mathcal{T}_4(\underline{\alpha})-\tilde{\mathcal{S}}(\underline{\alpha})) \right. \right.\nn \\
 &\phantom{+ e_d } -3 \mathcal{T}_1(\underline{\alpha})+ 3 \mathcal{T}_2(\underline{\alpha}) \bigg)  +2\mathds{P}\left[\mathcal{T}_{1}(\underline{\alpha})-\mathcal{T}_{2}(\underline{\alpha})-\mathcal{T}_{3}(\underline{\alpha})+\mathcal{T}_{4}(\underline{\alpha})\right]\bigg] \cdot \varepsilon^{\mu \nu \alpha \beta} p_{\nu} q_{\alpha} e_{\beta}^{(\lambda)} \; \mbox{ .}
\end{align}

Here  $\int Da = \int_0^1 {\rm d} \alpha_q \;\int_0^1 {\rm d}\alpha_{\qbar}\;\int_0^1 {\rm d} \alpha_{g} \; \delta(1-\alpha_q-\alpha_{\qbar}-\alpha_{g})$. The functions $\varphi$ and $\psi^{A}$, which are of twist 2 and twist 3, respectively, and $T_i$, $\mathcal{S}$, $\widetilde{\mathcal{S}}$, $\mathcal{S}_{\gamma}$, $T_{4}^{\gamma}$ and $\mathds{A}$, which have twist 4, are defined in Appendix \ref{A}. In order to simplify the expressions, several rather lengthy terms have been abbreviated via the function $\mathds{P}[x]$ defined in Appendix \ref{C}

The results for the diagrams Fig.\ref{diagrams}a,b reproduce those given in \cite{Aliev:1995wi}, if one takes into account the different nomenclature for the DAs and the different sign convention for the electromagnetic current. The contribution $T_{d}^{\mu}$ being rather small was not taken into account in \cite{Aliev:1995wi}.

In order to obtain a sum rule for $g_{D^*D\gamma}$, it is necessary to equate the hadronic representation Eq.\eqref{hadronic} and the twist-expansion, Eqs.(\ref{Ta},\ref{Tb},\ref{Tc},\ref{Td}), of the correlation function. This requires to control the contributions of excited states and of the continuum, that can be approximately evaluated using quark--hadron duality. The standard procedure involves a Borel transformation, which suppresses these contributions and a subsequent continuum subtraction.

As the momentum of the photon $q$ does not vanish in the real photon case $q^2=0$, we can treat $p_1$ and $p_2$ as independent, which allows a so-called double Borel transformation. This transformation can be carried out with the help of the formulas
\begin{align}
\label{borel1}
\mathcal{B}_{M_1^2}\mathcal{B}_{M_2^2}\left \lbrace\frac{\Gamma(\alpha)}{\left(m^2-\ubar p_1^2-u p_2^2 \right)^{\alpha}} \right \rbrace &= t^{2-\alpha} \delta\left(u-\frac{M_1^2}{M_1^2+M_2^2} \right)e^{-\frac{m^2}{t}}
\end{align}
and
\begin{align}
\label{borel2}
\mathcal{B}_{M_1^2}\mathcal{B}_{M_2^2}\left \lbrace\frac{1}{\left(m^2_{1}-p_1^2 \right)\left(m^2_{2}-p_2^2 \right)} \right \rbrace &= e^{-m_{1}^2/M_1^2-m_2^2/M_2^2}\mbox{ .}
\end{align}
Here $M_i^2$ is the Borel parameter corresponding to $p_i^2$ and $t=M_1^2 M_2^2/(M_1^2+M_2^2)$. As our final aim is a prediction for the branching ratio with help of the results from \cite{Belyaev:1994zk}, the method of how the continuum is subtracted must match the one employed therein. We will not go into the details of the procedure but instead would like to refer to the corresponding sections in \cite{Belyaev:1994zk} for a more detailed discussion. Furthermore, it should be noted that the non-trivial continuum subtraction of the perturbative contribution, Eq.\eqref{Ta}, is elaborated in \cite{Aliev:1995wi,Belyaev:1994zk}.

This leads to the following sum rule for $g_{D^*D\gamma}$:
\begin{align}
\label{SR}
g_{D^{*+}D^{+}\gamma} =& -\frac{ e^{\frac{M^2}{t}} }{f_{D^*}f_D m_D^2 m_{D^*} } \Bigg[ -\frac{m_c N_c}{4 \pi^2} \int_{m_c^2}^{S_0} \! {\rm d}s \left[ (e_d-e_c)\left(1-\frac{m_c^2}{s} \right)+e_c\ln\left(\frac{s}{m_c^2} \right) \right]e^{-S_{0}/t}\nn \\
& \phantom{-\frac{e^{\frac{M^2}{t}}} {f_{D^*}f_D m_D^2 m_{D^*}} }+\left( e^{-\frac{m_c^2}{t} } - e^{-\frac{S_0}{t}} \right) \left[ e_d \frac{ f_{3\gamma} m_c}{2}\phi^{A}(v) + e_d t \chi \qcond \varphi(v) \right.\nn \\
&\left. \phantom{-\frac{e^{\frac{M^2}{t}}}{f_{D^*}f_D m_D^2 m_{D^*}}}- e_d \frac{\mathds{A}(v)}{4} \qcond \left( 1 + \frac{m_c^2}{t} \right)  \right]
 +\mathcal{I}_F+\mathcal{I}_G \bigg]
\end{align}
with
\begin{align}
M^2=&\frac{2m_{D^*}^2 m_D^2}{m_D^2+m_{D^*}^2}\\
S_0=&\frac{2s_{D^*}^2 s_{D}^2}{s_{D^*}^2 +s_{D}^2}\\
v=&\frac{M_2^2}{M_2^2 + M_1^2} \mbox{ .}
\end{align}
$s_{D}$ and $s_{D^*}$ are the continuum thresholds for the $D^*$ and $D$, respectively. The functions $\mathcal{I}_F$ and $\mathcal{I}_G$  correspond to lengthy contributions of twist 4. The full expressions can be found in Appendix \ref{C}.
The decay width $\Gamma(D^{*+} \to D^+ \gamma)$ is then given by the formula \cite{Eletsky:1984qs}
\begin{align}
\label{width}
\Gamma(D^{*+} \to D^{+}\gamma)=\frac{g^2_{D^{*+} D^+\gamma} \alpha_{em}}{384 \pi^2}\frac{m^{2}_{D^*}-m^2_{D}}{m^{3}_{D^*}}\; \mbox{.}
\end{align}
The corresponding expressions for the $\Gamma(D^{*0} \to D^0 \gamma)$ decays can easily be found by replacing $e_d \leftrightarrow e_u$ in Eqs.(\ref{SR},\ref{width}).

\section{The branching ratios and the determination of $\chi(\mu=1{\rm GeV})$}
\label{3}

The branching ratio of the $ D^{*+} \to D^+ \gamma $ decay is defined as
\begin{align}
\mathcal{B} \left(D^{*+} \to D^+ \gamma \right)= \frac{\Gamma(D^{*+} \to D^{+} \gamma)}{\Gamma(D^{*+} \to D^{+} \gamma) + \Gamma(D^{*+} \to D^{+} \pi^0) + \Gamma(D^{*+} \to D^{0} \pi^{+})} \; \mbox{ .}
\end{align}
In order to determine $\mathcal{B} \left(D^{*+} \to D^+ \gamma \right)$ one needs the expressions for the decay widths of the $D^* \to D \pi$ decays. Up to twist 4 accuracy these have been determined in \cite{Belyaev:1994zk} \footnote{A more recent evaluation \cite{Khodjamirian:1999hb} also includes $\alpha_s$-corrections, for consistency we do not take these into account.}, which allows us to construct a sum rule for the branching ratio using Eq.\eqref{SR} and equations (44),(52) and (53) from Ref.\cite{Belyaev:1994zk}. A single sum rule for $\mathcal{B} \left(D^{*+} \to D^+ \gamma \right)$ has the distinct advantage that the dependence on several parameters that effect the individual sum rules for the widths will be significantly reduced. Most notably the dependence on the coupling constants $f_D$ and $f_{D^*}$, which are known only up to at best $10\%$, will vanish completely. It is, in principle, not possible to distentangle $\chi$ from $\varphi(v)$. Hence, we can only make a prediction for the product of the two.

As the mass difference $m_{D^*}-m_D$ is very small (of the order $0.07 m_{D^*}$) and can therefore be neglected, it is possible to use symmetric Borel parameters $M_1^2=M_2^2$, which corresponds to $v=1/2$. In the following analysis we will use the continuum threshold $S_0=6\;{\rm GeV}^2$ as determined from the sum rules for the couplings $f_D$ and $f_{D^*}$ \cite{Belyaev:1994zk}, a charm quark mass $m_c=1.3\pm 0.1\;{\rm GeV}$ and the Borel window $2\; {\rm GeV}^2 < t < 4 \;{\rm GeV}^2 $. The experimental value for the branching ratio \cite{PDG}
\begin{align}
\label{D+value}
\mathcal{B} \left(D^{*+} \to D^+ \gamma \right)=1.6 \pm 0.4 \%
\end{align}
will be an additional input parameter.
\begin{figure}
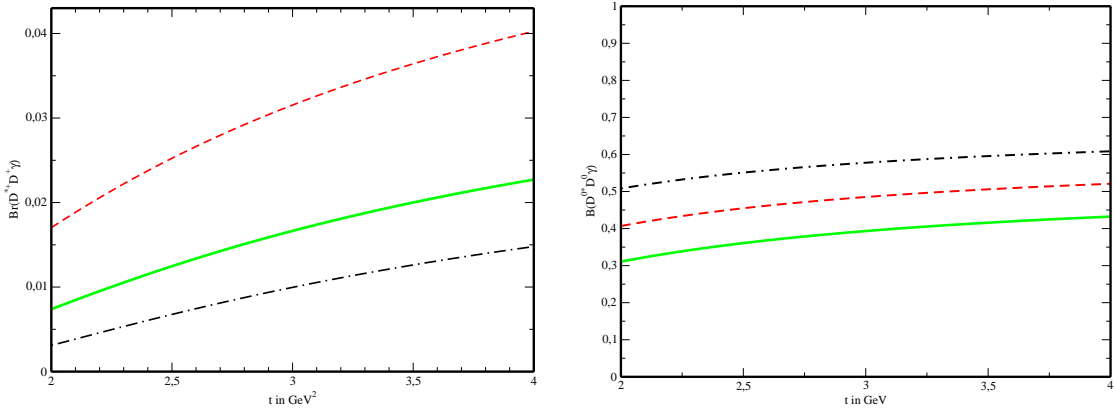

\label{BranchingratioPlot}
\epsfig{figure=BranchD+.eps, width=0.43 \textwidth} \;\;\;
\epsfig{figure=BranchD0.eps, width=0.43 \textwidth}
\caption{The left panel shows the plot of the branching ratio $\mathcal{B}(D^{*+} \to D^+\gamma)$ versus the Borel parameter $t$ for different values for the product $\chi(\mu)\varphi(1/2)$. The solid line represents the best fit to experiment and uses $\chi(\mu)\varphi(1/2)\approx 5.1\;{\rm GeV}^{-2}$. The dashed line corresponds to $\chi(\mu)\varphi(1/2)\approx 6.5\;{\rm GeV}^{-2}$ , the dashed-dotted to $\chi(\mu)\varphi(1/2)\approx 4.5\;{\rm GeV}^{-2}$. The right panel shows the plot of the branching ratio $\mathcal{B}(D^{*0} \to D^0\gamma)$. The best fit (solid line) yields $\chi(\mu)\varphi(1/2)\approx 3.5\;{\rm GeV}^{-2}$. The dashed and dashed-dotted lines correspond to $\chi(\mu)\varphi(1/2)\approx 4.8\;{\rm GeV}^{-2}$ and $\chi(\mu)\varphi(1/2)\approx 6.5\;{\rm GeV}^{-2}$, respectively.}
\end{figure}
In Fig. \ref{BranchingratioPlot} the plot of our sum rule for the branching ratio is shown using various choices for $\chi(\mu= 1.3{\rm GeV})\varphi(1/2)$. It can be seen, that branching ratio is indeed very sensitive to the value of this product. The best fit is achieved with
\begin{align}
\label{product1}
[\chi\varphi(1/2)](\mu= 1.3 {\rm GeV}^2) = 5.1^{+(0.4+0.3)}_{-(0.7+0.3)} \;{\rm GeV}^{-2}\mbox{ .}
\end{align}
The first given error is due to theoretical uncertainties, the second error stems from the experimental bounds.  \\

The branching ratio for the $D^{*0} \to D^0 \gamma$ decay $$\mathcal{B} \left(D^{*0} \to D^0 \gamma \right)= \frac{\Gamma(D^{*0} \to D^{0} \gamma)}{\Gamma(D^{*0} \to D^{0} \gamma) + \Gamma(D^{*0} \to D^{0} \pi)}$$
can be determined analogously and the corresponding plot can be found in Fig.\ref{BranchingratioPlot}. 
The sum rule has a weaker dependence on the fit parameter compared to the previous one. It should further be noted that the experimental uncertainties are significantly smaller.
A value of 
\begin{align}
\label{product2}
 [\chi\varphi(1/2)](\mu= 1.3 {\rm GeV}^2)= 3.5^{+(0.8+0.4)}_{-(0.6 + 0.3)} \;{\rm GeV}^{-2}
\end{align}
produces the best agreement with experiment \cite{PDG}
\begin{align}
 \mathcal{B} \left(D^{*0} \to D^0 \gamma \right)= 38.1 \pm 2.9 \% \; \mbox{ .}
\end{align}
The identification of the errors follows Eq.\eqref{product1}. It should be noted that the result in Eq.\eqref{product2} shows a rather strong dependence on the charm quark mass, which is one reason for the rather large relative errors. However, this is not the case for Eq.\eqref{product1} as the sum rule for $\mathcal{B} \left(D^{*+} \to D^+ \gamma \right)$ is almost insensitive to the $c$--quark mass.

 In order get an expression for $\chi$ alone one has to introduce an explicit value for $\varphi(1/2)$. The natural choice is to use the so called asymptotic form $\varphi(u)=6 u \ubar$, see discussion in \cite{Ball:2002ps}. Taking into account the scale dependence of the product $\chi \varphi$,  one gets 
\begin{align}
\chi(\mu=1{\rm GeV}^2)= 3.5^{+0.5}_{-0.7} \;{\rm GeV}^{-2}
\end{align}
from the results for $D^{*+} \to D^+ \gamma$ and 
\begin{align}
\chi(\mu=1{\rm GeV}^2)=  2.4^{+0.9}_{-0.7} \;{\rm GeV}^{-2},
\end{align}
form $D^{*0} \to D^0 \gamma$.

\begin{figure}
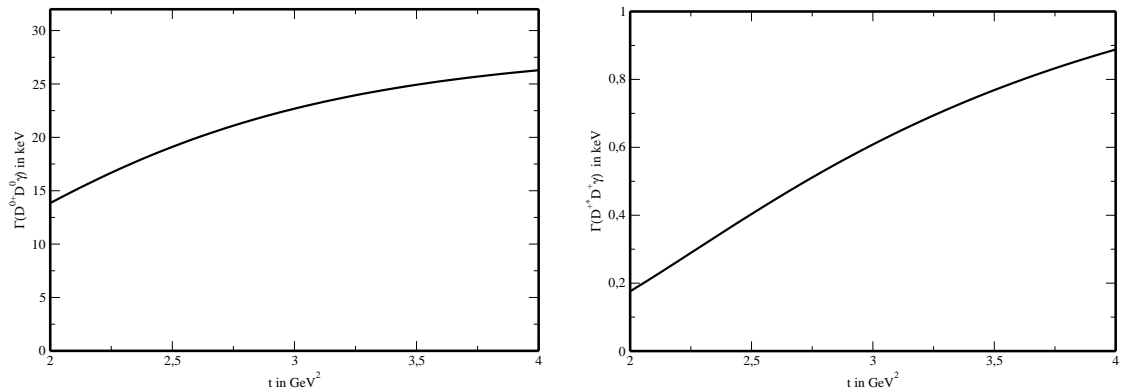

\epsfig{figure=WidthD0.eps, width=0.43 \textwidth} \;\;\;
\epsfig{figure=WidthD+.eps, width=0.43 \textwidth}
\caption{The sum rules for the radiative $D^*$ decay widths using $\chi \psi(1/2)=4.65 \;{\rm GeV}^{-2}$. \label{fig4}}
\end{figure}

For completeness, we also determine the decay widths for the different radiative $D$-decays. In this case on has to use a fixed value for $\chi(\mu=1.3 {\rm GeV})\varphi(1/2)$, we use $4.73\pm 0.45GeV^{-2}$, which corresponds to $\varphi(1/2)=3/2$ and $\chi(\mu=1{\rm GeV}^2)=3.15 \pm 0.3 {\rm GeV}^{-2}$. In this case, the coupling constants $f_D$ and $f_{D^*}$ are relevant and we will use the sum rules given in Eqs.(46,47) in \cite{Belyaev:1994zk}. The results are shown in  Fig.\ref{fig4}.
\begin{align}
\label{WidthPlots}
\Gamma(D^{*0} \to D^0 \gamma)= 20 \pm 6 \; {\rm keV} \\
\Gamma(D^{*+} \to D^+ \gamma)= 0.55 \pm 0.3\;  {\rm keV}
\end{align}
The large uncertainties are mainly due to the uncertainties in the coupling constants $f_D$ and $f_{D^*}$, that are not present in the sum rules for the branching ratios. A comparison with the CLEO data \cite{Bartelt:1997yu} $\Gamma(D^{*+})_{\rm total} = 96 \pm 26 \; {\rm keV}$ would lead to a branching ratio of $\approx 0.6 \%$, which would be below the value given in \eqref{D+value}. While the experimental number has yet to be determined more precisely, it seems to indicate that QCD sum rules generally underestimate the decay widths in $D^* \to D \gamma$ and  $D^* \to D \pi$, see also \cite{Khodjamirian:1999hb, Aliev:1995wi, Eletsky:1984qs}. However, the LCSR predictions for the branching ratios, that are measured more precisely then the widths, typically agree rather well with experiment.

\section{Conclusions}
\label{4}

We have calculated the radiative decay constants of the $D^*$ using the approach of light-cone sum rules. This enabled us to use results from \cite{Belyaev:1994zk} to construct a sum rule for the branching ratios $\mathcal{B}(D^{*+} \to D^+\gamma)$ and $\mathcal{B}(D^{*0} \to D^0\gamma)$. As this sum rule is dominated by terms of twist 2, that are proportional to the magnetic susceptibility of the quark consdensate $\chi$, it is possible to determine its numerical value by fitting the sum rule to the experimentally determined value of the branching ratio. However, the magnetic susceptibility cannot be accessed directly as it is "masked" by an additional factor $\varphi(1/2)$, and only the product of the two can be fitted.

Taking the naive average of Eq.\eqref{product1} and Eq.\eqref{product2}, the best agreement with experimental data can be achieved by the choice
\begin{align}
[\chi \varphi(1/2)](\mu=1.3 {\rm GeV}) \approx 4.3 \pm 0.7 \; {\rm GeV}^{-2}.
\end{align}
 In order to be able to compare our result to other calculations of $\chi$, it is necessary to assume an explicit value for $\varphi(1/2)$. The choice of the asymptotic wave function, which has produced good agreement with experiment in the past, leads to 
\begin{align}
\chi(\mu=1{\rm GeV})= 2.85 \pm 0.5 {\rm GeV}^{-2}.
\end{align}
This value is below the most recent QCD sum rule result \cite{Ball:2002ps} and is still affected by rather large uncertainties, which are due to the limited precision of both sum rule parameters and the experimental data. However, this result is very close to the one obtained in \cite{Dorokhov:2005pg} using an instanton liquid model.
Alternatively, assuming the standard value $\chi(\mu=1\;{\rm GeV})=3.15 \;{\rm GeV}^{-2}$ the estimate 
\begin{align}
\varphi(1/2, \mu=1 \;{\rm GeV}) \approx 1.35 \pm 0.3
\end{align}
can be obtained. At first glance, this seems to indicate that the shape of the wave function may be more flat compared the the asymptotic form, however, our calculation is not yet precise enough to support such a statement. 

\section*{Acknowledgements}
I am very grateful to V. Braun for clarifying discussions, to A. Khodjamirian for enlightening comments and to A. Lenz for proofreading the manuscript. 
\newpage

\begin{appendix}
\section{Photon distribution amplitudes}
\label{A}

For completeness we collect the relevant photon distribution amplitudes for the $D^*\to D \gamma$ transition according to \cite{Ball:2002ps}. Note that in \cite{Ball:2002ps} the photon momentum has the opposite sign and the parameterization of the separation of antiquark and quark is different.

The path-ordered exponents
\begin{equation*}
\left[ x,y \right]={\rm Pexp}\left \lbrace i \int \! dt \;(x-y)_{\mu} \left[ e_q A^{\mu}(tx-\overline{t}y) + g B^{\mu}(tx-\overline{t}y) \right] \right \rbrace
\end{equation*}
assure gauge invariance of the matrix elements. It is important, that the electromagnetic potential $A^{\mu}$ is included in addition to the gluon potential$B^{\mu}$, as additional terms to those given in \cite{Ball:2002ps} will occur otherwise.

\subsection{Twist-2 and Twist-4 DAs}
The leading-twist DA reads
 \begin{align}
\label{SigmaMatrixelement}
\bra{0} \qbar(0) \left[ 0,x \right]\sigma_{\alpha \beta} q(x) \ket{0}_F&=e_q \left\langle \qbar q \right\rangle \int _0 ^1  \! du \; \chi \varphi (u) F_{\alpha \beta}(ux) \nn \\
     & \phantom{=} +\frac{e_q \left\langle \qbar q \right\rangle}{16} \int_0^1 \!du\; x^2 \mathds{A}(u) F_{\alpha \beta}(ux) \nn \\
     & \phantom{=} +\frac{e_q \left\langle\qbar q \right\rangle}{8}\int_0^1 \!du\;\mathds{B}(u)x^{\rho}\left(x_{\beta}F_{\alpha \rho}(ux) -x_{\alpha}F_{\beta \rho}(ux) \right)
\end{align}
with 
\begin{align}
\varphi(u)&=\varphi^{{\rm asy.}}(u)=6 u(1-u) \\
\mathds{A}(u)&=40 u (1-u)\left(3\kappa -\kappa^{+}+1 \right)+8\left(\zeta^{+}_2-3\zeta_2 \right) \times 
\nn \\
 &\phantom{=} \left[ u(1-u)\left( 2+13u(1-u) \right)+2u^3\left(10-15u+6u^2 \right)\ln (u)\right.
\nn \\
&\phantom{=} \left.+ 2(1-u)^3\left(10-15(1-u)+6{1-u}^2 \right)\ln (1-u)\right]
 \\
\mathds{B}(u)&=40 \int_0^u \! d\alpha \; (u-\alpha) \left(1+3 \kappa^+ \right)\left[ -\frac{1}{2}+\frac{3}{2} (2\alpha -1)^2 \right] \mbox{ .}
\end{align}

\begin{align}
\label{photonDA1}
\bra{0}& \qbar(0) e_q \left[ 0,x \right] F_{\mu \nu} (ux) q(x) \ket{0}_F=e_q \qcond \int \Da \mathcal{S}_{\gamma}(\underline{\alpha})F_{\mu \nu}(\alpha_u x)\\
\label{photonDA2}
\bra{0}& \qbar(0) e_q \left[ 0,x \right] \sigma _{\alpha \beta} F_{\mu \nu} (ux) q(x) \ket{0}_F=
\nn\\
&-\frac{e_q \qcond}{qx} \left[ q_{\alpha} q_{\mu} e^{\lambda}_{\bot \nu} x_{\beta}-  q_{\beta} q_{\mu} e^{\lambda}_{\bot \nu}- q_{\alpha} q_{\nu} e^{\lambda}_{\bot \mu} x_{\beta} +q_{\beta} q_{\nu} e^{\lambda}_{\bot \mu} x_{\alpha} \right]\mathcal{T}^{\gamma}_4(u,qx)\\
\label{DistAmpl1}
\bra{0}&\qbar(0) \left[ 0,ux \right]g G_{\mu \nu }(ux)\left[ ux,x \right]q(x)\ket{0}_F=e_q \qcond \int\!\! \Da \mathcal{S}(\underline{\alpha}) F_{\mu \nu}(\alpha_u x)\\
\label{DistAmpl2}
\bra{0}&\qbar(0) \left[ 0,ux \right]i\gamma_5 g \widetilde{G}^{\mu \nu }(ux) \left[ ux,x \right] q(x)\ket{0}_F=e_q \qcond \int\!\! \Da \mathcal{\widetilde{S}}(\underline{\alpha}) F_{\mu\nu}(\alpha_u x)\\
\label{DistAmpl5}
\bra{0}&\qbar(0)\left[ 0,ux \right]\sigma_{\alpha \beta} g G_{\mu \nu}(ux) \left[ ux,x \right]q(x)\ket{0}_F=
\nn \\
 &= - e_q \qcond \left[ q_{\alpha} e ^{(\lambda)}_{\bot \mu}g_{\beta \nu}^{\bot}-q_{\beta}e ^{(\lambda)}_{\bot \mu}g_{\alpha \nu}^{\bot}-q_{\alpha}e ^{(\lambda)}_{\bot \nu}g_{\beta \mu}^{\bot}+q_{\beta}e ^{(\lambda)}_{\bot \nu}g_{\alpha \mu}^{\bot} \right] \mathcal{T}_1(u,qx)\quad\;\;\!
\nn\\
 & \phantom{=}- e_q \qcond \left[ q_{\mu} e ^{(\lambda)}_{\bot \alpha}g_{\beta \nu}^{\bot}-q_{\mu}e ^{(\lambda)}_{\bot \beta}g_{\alpha \nu}^{\bot}-q_{\nu}e ^{(\lambda)}_{\bot \alpha}g_{\beta \mu}^{\bot}+q_{\nu}e ^{(\lambda)}_{\bot \beta}g_{\alpha \mu}^{\bot} \right]\mathcal{T}_2(u,qx)\quad\;\;\!
\nn\\
 &\phantom{=}-\frac{e_q \qcond}{qx} \left[ q_{\alpha} q_{\mu} e^{\lambda}_{\bot \beta} x_{\nu}-  q_{\beta} q_{\mu} e^{\lambda}_{\bot \alpha}- q_{\alpha} q_{\nu} e^{\lambda}_{\bot \beta} x_{\mu} +q_{\beta} q_{\nu} e^{\lambda}_{\bot \alpha} x_{\mu} \right]\mathcal{T}_3(u,qx)
\nn\\
&\phantom{=}-\frac{e_q \qcond}{qx} \left[ q_{\alpha} q_{\mu} e^{\lambda}_{\bot \nu} x_{\beta}-  q_{\beta} q_{\mu} e^{\lambda}_{\bot \nu}- q_{\alpha} q_{\nu} e^{\lambda}_{\bot \mu} x_{\beta} +q_{\beta} q_{\nu} e^{\lambda}_{\bot \mu} x_{\alpha} \right]\mathcal{T}_4(u,qx)
\end{align}
Here we used 
\begin{eqnarray}
\int \Da &=& \int_0^1 \!d\alpha_q \; \int_0^1 \!d\alpha_{\qbar} \; \int_0^1 \!d\alpha_g \; \delta(1-\alpha_q-\alpha_{\qbar}-\alpha_g) \\
\alpha_u&=&\alpha_q+u\alpha_g\\
g_{\mu \nu}^{\perp}&=&g_{\mu \nu}-\frac{q_{\mu}x_{\nu}+q_{\nu}x_{\mu}}{qx}\\
e^{\perp\;\!(\lambda)}_{\mu}&=&g_{\mu \nu}^{\perp}\;\!e^{\nu\;\!(\lambda)}
\end{eqnarray}
and
\begin{align}
\mathcal{S}(\underline{\alpha})&=30\alpha_g^2 \left[ \left( \kappa +\kappa^+ \right)\left(1-\alpha_g \right)+ \left( \zeta_1+\zeta_1^+ \right)\left( 1-\alpha_g \right) \left( 1-2\alpha_g \right) \right.
\nn \\
&\phantom{=}\left.+\zeta_2 \left( 3\left( \alpha_{\qbar}-\alpha_q \right)^2-\alpha_g \left( 1-\alpha_g \right) \right) \right]
\\
\widetilde{\mathcal{S}} (\underline{\alpha})&=-30\alpha_g^2 \left[ \left( \kappa -\kappa^+ \right)\left(1-\alpha_g \right)+ \left( \zeta_1-\zeta_1^+ \right) \left( 1-\alpha_g \right) \left( 1-2\alpha_g \right)\right.
\nn \\
&\phantom{=}\left.+\zeta_2\left( 3\left( \alpha_{\qbar}-\alpha_q \right)^2-\alpha_g\left(1-\alpha_g \right) \right) \right]
\\
S_{\gamma} \left( \underline{\alpha} \right) &=60 \alpha_g^2\left( \alpha_{\qbar} + \alpha_q \right)\left( 4 - 7 \left( \alpha_q + \alpha_{\qbar} \right) \right) 
\\
\mathcal{T}_i(u,qx)&=\int \Da \; e^{i\alpha_u qx} T_i(\underline{\alpha})
\end{align}
with
\begin{align}
T_1(\underline{\alpha})&=-120 \left( 3 \zeta_2 +\zeta_2^+ \right)\left( \alpha_{\qbar}-\alpha_q \right) \alpha_{\qbar}\alpha_q \alpha_g
\\
T_2(\underline{\alpha})&=30 \alpha_g^2\left(\alpha_{\qbar}-\alpha_q \right) \left[\left(\kappa -\kappa^+ \right)+\left(\zeta_1-\zeta_1^+ \right)\left( 1-2\alpha_g \right) +\zeta_2 \left( 3-4 \alpha_g \right) \right]
\\
T_3(\underline{\alpha})&=-120 \left( 3 \zeta_2 -\zeta_2^+ \right)\left( \alpha_{\qbar}-\alpha_q \right) \alpha_{\qbar}\alpha_q \alpha_g
\\
T_4(\underline{\alpha})&=30 \alpha_g^2\left(\alpha_{\qbar}-\alpha_q \right) \left[\left(\kappa +\kappa^+ \right)+\left(\zeta_1+\zeta_1^+ \right)\left( 1-2\alpha_g \right) +\zeta_2 \left( 3-4 \alpha_g \right) \right]
\\
T^{\gamma}_4(\underline{\alpha})&=60 \alpha_g^2\left( \alpha_{\qbar}-\alpha_q \right)\left( 4 - 7 \left( \alpha_q + \alpha_{\qbar} \right) \right) \mbox{ .}
\end{align}
The abbreviation $\underline{\alpha}$ represents $(\alpha_q, \alpha_{\qbar}, \alpha_g)$. The values of the various constants can be found in table \ref{Numbers}.

It should be noted that the matrix element 
$$\bra{0} \qbar(0) e_q \left[ 0,x \right] \sigma _{\alpha \beta} F_{\mu \nu}(ux) q(x) \ket{0}_F$$ 
vanishes exactly if one sums up the whole conformal expansion. The expansion itself has, however, non-zero coefficients and thus in next-to-leading order in conformal spin the matrix element is different from zero. For the same reason the matrix element
$$ \bra{0} \qbar(0) e_q \left[ 0,x \right] F_{\mu \nu} (ux) q(x) \ket{0}_F $$
has herein mentioned form and not $e_q \qcond F_{\mu \nu} (ux) \mbox{ .}$

\subsection{Twist-3 DAs}

\begin{align}
\label{Twist3MatrixelementVektor}
\bra{0} \qbar(0)\left[ 0,x \right] \gamma_{\alpha} q(x) \ket{0}_F =-\frac{e_q}{2}f_{3 \gamma} \int_0 ^1\! du \; \overline{\psi}^{(V)}(u) x^{\rho}F_{\rho \alpha} 
\\
\label{Twist3MatrixelementAxial}
\bra{0} \qbar(0) \left[ 0,x \right]\gamma_{\alpha} \gamma_5 q(x) \ket{0}_F =-i\frac{e_q}{4}f_{3 \gamma} \int_0 ^1\! du \; \psi^{(A)}(u) x^{\rho}\widetilde{F}_{\rho \alpha} 
\\
\label{DistAmpl3}
\bra{0}\qbar(0) \left[ 0,ux \right]i g \gamma_{\alpha} G_{\mu \nu }(ux)\left[ ux,x \right]q(x)\ket{0}_F= \qquad\qquad\qquad
\nn \\
 =e_q f_{3\gamma}q_{\alpha} \left[q^{\nu} e^{(\lambda)}_{\bot \mu} - q^{\mu} e^{(\lambda)}_{\bot \nu}  \right] \int \Da \mathcal{V}(\underline{\alpha}) e^{i\alpha_uqx}\quad\\
\label{DistAmpl4}
\bra{0}\qbar(0) \left[ 0,ux \right] g \gamma_{\alpha}\gamma_5 \widetilde{G}_{\mu \nu }(ux)\left[ ux,x \right]q(x)\ket{0}_F=\qquad\qquad\qquad
\nn \\
=e_q f_{3\gamma}q_{\alpha} \left[q^{\nu} e^{(\lambda)}_{\bot \mu} - q^{\mu} e^{(\lambda)}_{\bot \nu}  \right] \int \Da \mathcal{A}(\underline{\alpha}) e^{i\alpha_uqx}\quad
\end{align}
Where
\begin{align}
\overline{\psi}^{(V)}(u)&=-20u(1-u)(2u-1)
\nn \\ 
&\phantom{=} +\frac{15}{16} \left(\omega_{\gamma}^A-3\omega_{\gamma}^{V} \right)u (1-u) (2u-1)\left(7(2u-1)^2-3 \right)\\
{\psi}^{(A)}(u)&=(1-(2u-1)^2) \left( 5 \left( 2u-1 \right)^2 -1\right)\frac{5}{2} \left( 1+\frac{9}{16}\omega_{\gamma}^{V}-\frac{3}{16} \omega_{\gamma}^{A}\right)\\
\mathcal{V}(\underline{\alpha})&=540 \omega_{\gamma}^V \left( \alpha_q - \alpha_{\qbar} \right) \alpha_q \alpha_{\qbar} \alpha_{g}^2\\
\mathcal{A}(\underline{\alpha})&=360\alpha_q \alpha_{\qbar} \alpha_g^2 \left[1+\omega_{\gamma}^A\frac{1}{2} \left( 7 \alpha_g-3 \right) \right] \mbox{ .}
\end{align}
\newpage
\subsection{Numerical values for the parameters at the renormalization scale $\mu=1{\rm GeV}$}

\begin{table}[h]
\begin{center}
\begin{tabular}{c||c}
\hline
$\chi$ & $3.15\pm 0.3\;\!{\rm GeV}^{-2}$  \\\hline
$\kappa$ &  $0.15 $ \\\hline
$\kappa^+$ & $-0.05  $ \\\hline
$\zeta_1$ & $0.4 $ \\\hline
$\zeta_1^+$ &$0  $ \\\hline
$\zeta_2$ & $0.3 $ \\\hline
$\zeta_2^+$ &$0  $ \\\hline
$f_{3\gamma}$ & $-(4\pm 2)\cdot 10^{-3}\;\!{\rm GeV}^2$      \\\hline
$\omega_{\gamma}^A$ &$-2.1\pm 1.0 $ \\\hline
$\omega_{\gamma}^V$ & $3.8\pm 1.8 $\\\hline
$\qcond$ &$-(240 \pm 10\;\!{\rm MeV} )^3$ \\\hline
\end{tabular}
\end{center}

\caption{Numerical values and uncertainties of the relevant parameters \cite{Ball:2002ps,Ball:2007zt, Balitsky:1989ry} \label{Numbers}.} 
\end{table}

\section{The functions $\mathds{P}$ and $\mathcal{I}_{F}, \mathcal{I}_{G}$ }
\label{C}

In this Section we have gathered the explicit expressions for the three functions $\mathds{P}$, $\mathcal{I}_F$ and $\mathcal{I}_G$ that appear in Eqs.(\ref{Tc},\ref{Td}) and (\ref{SR}).

\begin{align}
\mathds{P}[X(\underline{\alpha})]=&8 e_c \qcond \int_0^1\! {\rm d}u \int_0^1\! {\rm d}\alpha_q \int_0^{\alpha_q}\! {\rm d}\alpha_q' \int_0^{1-\alpha_q'}\! {\rm d} \alpha_{\qbar}  u\frac{pq X(\alpha_q,\alpha_{\qbar},1-\alpha_q-\alpha_{\qbar})}{m_c^2-(p+(u-u \alpha_q+\ubar \alpha_{\qbar})q)^2}   \nn \\
&\phantom{8 e_c \qcond}-\int_0^1\! {\rm d}u \int_0^1\! {\rm d}\alpha_q \int_0^{1-\alpha_q}\! {\rm d}\alpha_{\qbar} \int_0^{\alpha_{\qbar}}\! {\rm d}\alpha_{\qbar}'  \ubar \frac{pq X(\alpha_q,\alpha_{\qbar}',1-\alpha_q-\alpha_{\qbar}')}{m_c^2-(p+ \ubar \alpha_{\qbar} q)^2} \nn \\
&\phantom{8 e_c \qcond}+\int_0^1\! {\rm d}u \int_0^1\! {\rm d}\alpha_q \int_0^{\alpha_q}\! {\rm d}\alpha_{q}' \int_0^{1-\alpha_{q}'}\! {\rm d}\alpha_{\qbar}'  \ubar \frac{pq X(\alpha_q',\alpha_{\qbar},1-\alpha_q'-\alpha_{\qbar})}{m_c^2-(p+\ubar(1- \alpha_{q})q)^2} 
\\
\mathcal{I}_{F}=&\phantom{-} e_c\qcond \int_{0}^{1/2}\!\!\!{\rm d}\alpha_q \int_{0}^{1/2}\!\!\!\!{\rm d}\alpha_{\qbar}\frac{1}{1 - \alpha_q-  \alpha_{\qbar}}\left(e^{-\frac{m_c^2}{t}}-e^{-\frac{S_0}{t}}\right) \mathcal{S}_{\gamma}(\alpha_q, \alpha_{\qbar}, 1-\alpha_{q}-\alpha_{\qbar})\nn \\
&-e_c\qcond \int_{0}^{1/2}\!\!\!\!{\rm d}\alpha_q \int_{0}^{1/2}\!\!\!\!{\rm d}\alpha_{\qbar} \frac{2-\alpha_q-3\alpha_{\qbar}}{(1 - \alpha_q-  \alpha_{\qbar})^2} \left(e^{-\frac{m_c^2}{t}}-e^{-\frac{S_0}{t}}\right) \mathcal{T}_{\gamma}^4(\alpha_q, \alpha_{\qbar}, 1-\alpha_{q}-\alpha_{\qbar} ) \nn \\
& +2 e_c\qcond \left(e^{-\frac{m_c^2}{t}}-e^{-\frac{S_0}{t}}\right) \frac{m_D^2-m_{D^*}^2}{t}\nn \\
&\phantom{2 e_c\qcond} \times \bigg\lbrace \int_0^{1/2} \! {\rm d}\alpha_q \int_0^{\alpha_q} \! {\rm d}\alpha_q' \int_0^{1/2} \! {\rm d}\alpha_{\qbar} \frac{1/2-\alpha_{\qbar}}{(1-\alpha_q-\alpha_{\qbar})^2}  \mathcal{T}_{\gamma}^4(\alpha_q, \alpha_{\qbar}, 1-\alpha_{q}-\alpha_{\qbar})\nn \\
&\phantom{2 e_c\qcond \times \bigg\lbrace}+\int_{1/2}^{1} \! {\rm d}\alpha_q \int_0^{1/2} \! {\rm d}\alpha_q' \int_{1/2}^{1-\alpha_q} \! {\rm d}\alpha_{\qbar} \frac{1/2-\alpha_{\qbar}} {(1-\alpha_q-\alpha_{\qbar})^2}  \mathcal{T}_{\gamma}^4 (\alpha_q, \alpha_{\qbar}, 1-\alpha_{q}-\alpha_{\qbar})\nn \\
&\phantom{2e_c \qcond \times \bigg \lbrace} -\int_{0}^{1/2} \! {\rm d}\alpha_q \int_0^{1-\alpha_q} \! {\rm d}\alpha_{\qbar} \frac{1}{2 \alpha_{\qbar}^2} \int_{0}^{\alpha_{\qbar}} \! {\rm d}\alpha_{\qbar}' \mathcal{T}_{\gamma}^4(\alpha_q, \alpha_{\qbar}', 1-\alpha_{q}-\alpha_{\qbar}')\nn \\
&\phantom{2 e_c\qcond \times \bigg\lbrace}+\int_{0}^{1/2} \! {\rm d}\alpha_q \frac{1}{2(1- \alpha_{q})^2} \int_0^{\alpha_q} \! {\rm d}\alpha_{q}' \int_{0}^{1-\alpha_{q}'} \! {\rm d}\alpha_{\qbar} \mathcal{T}_{\gamma}^4(\alpha_q', \alpha_{\qbar}, 1-\alpha_{q}'-\alpha_{\qbar}) \bigg \rbrace
\\
\mathcal{I}_{G}=&\phantom{-} e_q\qcond \int_{0}^{1/2}\!\!\!\!{\rm d}\alpha_q \int_{0}^{1/2}\!\!\!\!\!{\rm d}\alpha_{\qbar}\! \left(e^{-\frac{m_c^2}{t}}-e^{-\frac{S_0}{t}}\right) 
\left[ \frac{1}{1 - \alpha_q-  \alpha_{\qbar}} \mathcal{S} +\frac{2-3\alpha_{\qbar}-\alpha_q}{(1-\alpha_q-\alpha_{\qbar})^2}\! (\mathcal{T}_3-\mathcal{T}_4) \right.\nn \\ 
&\phantom{- e_q\qcond \int_{0}^{1/2} }\;\;\;\left.-\frac{3}{1 - \alpha_q-  \alpha_{\qbar}}(\mathcal{T}_1-\mathcal{T}_2)-\frac{\alpha_{\qbar}-\alpha_q}{(1 - \alpha_q-  \alpha_{\qbar})^2} \tilde{\mathcal{S}}\right] (\alpha_q, \alpha_{\qbar}, 1-\alpha_{q}-\alpha_{\qbar})\nn \\
& +2 e_q \qcond \left(e^{-\frac{m_c^2}{t}}-e^{-\frac{S_0}{t}}\right) \frac{m_D^2-m_{D^*}^2}{t}\nn \\
&\phantom{+2} \times \bigg\lbrace \int_0^{1/2} \!\!\!\! {\rm d}\alpha_q \int_0^{\alpha_q} \!\!\!\! {\rm d}\alpha_q' \int_0^{1/2} \!\!\!\! {\rm d}\alpha_{\qbar} \frac{1/2-\alpha_{\qbar}}{(1-\alpha_q-\alpha_{\qbar})^2} \left[\mathcal{T}_1-\mathcal{T}_2-\mathcal{T}_3+\mathcal{T}_4 \right](\alpha_q, \alpha_{\qbar}, 1-\alpha_{q}-\alpha_{\qbar})\nn \\
&\phantom{+2 \times \bigg\lbrace}+\int_{1/2}^{1} \!\!\!\! {\rm d}\alpha_q \int_0^{1/2} \!\!\!\! {\rm d}\alpha_q' \int_{1/2}^{1-\alpha_q} \!\!\!\! {\rm d}\alpha_{\qbar} \frac{1/2-\alpha_{\qbar}} {(1-\alpha_q-\alpha_{\qbar})^2}  \left[\mathcal{T}_1-\mathcal{T}_2-\mathcal{T}_3+\mathcal{T}_4 \right] (\alpha_q, \alpha_{\qbar}, 1-\alpha_{q}-\alpha_{\qbar})\nn \\
&\phantom{+2 \times \bigg \lbrace} -\int_{0}^{1/2} \!\!\!\! {\rm d}\alpha_q \int_0^{1-\alpha_q} \!\!\!\! {\rm d}\alpha_{\qbar} \frac{1}{2 \alpha_{\qbar}^2} \int_{0}^{\alpha_{\qbar}}\!\!\!\! {\rm d}\alpha_{\qbar}' \left[\mathcal{T}_1-\mathcal{T}_2-\mathcal{T}_3+\mathcal{T}_4 \right](\alpha_q, \alpha_{\qbar}', 1-\alpha_{q}-\alpha_{\qbar}')\nn \\
&\phantom{+2 \times \bigg\lbrace}+\int_{0}^{1/2} \!\!\!\! {\rm d}\alpha_q \frac{1}{2(1- \alpha_{q})^2} \int_0^{\alpha_q}\!\!\!\! {\rm d}\alpha_{q}' \int_{0}^{1-\alpha_{q}'} \!\!\!\! {\rm d}\alpha_{\qbar} \left[\mathcal{T}_1-\mathcal{T}_2-\mathcal{T}_3+\mathcal{T}_4 \right](\alpha_q', \alpha_{\qbar}, 1-\alpha_{q}'-\alpha_{\qbar}) \bigg \rbrace
\end{align}

\end{appendix}

\end{document}